\documentclass[superscriptaddress,aps,prd,twocolumn]{revtex4-2}
\usepackage[english]{babel}
\usepackage[utf8]{inputenc}
\usepackage{amsmath,ulem}
\usepackage{amssymb}
\usepackage{graphicx}
\usepackage{color}
\usepackage{float}
\usepackage{slashed}
\usepackage{lipsum}

\usepackage{hyperref}
\usepackage[utf8]{inputenc}

\newcommand{\mf}[1]{{\bf \textcolor[rgb]{0.59, 0.29, 0.0}{#1}}}

\definecolor{green}{RGB}{0,200,0}

\begin{document}

\title{Determination of the symmetry energy from the neutron star equation of state}

\author{Pedro Barata de Tovar}
\email{pbtcs2@gmail.com}
\affiliation{CFisUC, Department of Physics, University of Coimbra, 
	P-3004 - 516  Coimbra, Portugal}
	
\author{Márcio Ferreira}
\email{marcio.ferreira@uc.pt}
\affiliation{CFisUC, 
	Department of Physics, University of Coimbra, P-3004 - 516  Coimbra, Portugal}
	
\author{Constança Providência}
\email{cp@fis.uc.pt}
\affiliation{CFisUC, 
	Department of Physics, University of Coimbra, P-3004 - 516  Coimbra, Portugal}

\date{\today}

\begin{abstract}
We analyze the uncertainties introduced in the determination of the neutron star matter proton fraction, in a range of densities close to the saturation density, if the cold $\beta$-equilibrium neutron star matter equation of state (EOS) is known. In particular, we discuss the effect of neglecting the muon contribution and of  considering that the energy density of nuclear matter is well described by taking only terms until second order in the proton-neutron asymmetry. 
It is shown that two types of uncertainties may be associated with the extraction of the  symmetry energy from the $\beta$-equilibrium equation of state: 
an overestimation   if    terms  above  the parabolic  approximation  on  the  asymmetry  parameter are neglected,  or an  underestimation if the muon contribution is not considered.
The effect of the uncertainty on the symmetric nuclear matter EOS  on the determination of the proton fraction  is discussed. It could be shown that the neutron star mass-radius curve is sensitive to the parabolic approximation on the asymmetry parameter.
\end{abstract}

\maketitle

\section{Introduction}
\label{introduction}

Neutron stars (NS) are a real laboratory to determine the equation of state (EOS) of cold nuclear matter under extreme conditions, in particular, dense and very neutron rich nuclear matter. These conditions are not presently attainable with terrestrial experiments.  However, the observation of different signals from the Universe is setting  a large number of constraints which will allow the determination of the high density EOS of nuclear matter.  These observations include:  the detection of two solar mass pulsars such as the pulsars PSR J1614$-$2230 with $M=1.908\pm0.016\, M_\odot$
\cite{Arzoumanian2017,Fonseca2021,Demorest2010}, PSR J0348$+$0432 with $M=2.01 \pm 0.04 M_\odot$
\cite{Antoniadis2013} or MSP J0740$+$6620,
with a mass 
$2.08^{+0.07}_{-0.07}M_\odot$ 
\cite{Cromartie2019,Fonseca2021}, and the radius $12.39^{+1.30}_{-0.98}$ km determined  from the joint analysis of the Neutron Star
Interior Composition Explorer (NICER) and XMM-Newton data  \cite{Riley2021} (see also \cite{Miller2021}); the estimations of the mass and the radius of the
pulsar PSR J0030+0451 by  NICER 
\cite{Riley_2019,Miller_2019}; the detection  of the gravitational wave  GW170817 of a binary NS merger   by the LIGO-Virgo collaboration (LVC)  \cite{GW170817,Abbott2018PRL} or the detection of  GW190425 \cite{GW190425}.
 The GW170817 event was followed by electromagnetic
counterparts,  the gamma-ray burst (GRB) GRB170817A \cite{grb}, and the
electromagnetic transient AT2017gfo \cite{kilo}. It has been shown that these observations also set constraints on the tidal deformability, in particular,  the lowest limit of the tidal deformability of  a 1.37$M_\odot$
star was set in different studies to  $\Lambda_{1.37M_\odot} > 210$ \cite{Bauswein2019}, 300 \cite{Radice2018},
279 \cite{Coughlin2018}, and 309 \cite{Wang2018}. 

In the present study, we will discuss which information  can be extracted from the knowledge of the $\beta$-equilibrium EOS of NS matter. The determination of the NS EOS is in principle possible once information on the mass and the radius of several NS is obtained \cite{Lindblom1992,Steiner2010,Fujimoto:2017cdo,Fujimoto:2019hxv}. More measurements of NS  radii by NICER are expected  in the near future. Also many more precise measurements of NS masses are expected when the Square Kilometer Array enters in its second phase \cite{Watts2015}.

The determination of the nuclear  symmetry energy in a wide range of densities has been a topic of research during the last decade by many authors, see the Topical Issue \cite{VidanaEPJA} or the recent review \cite{Li2021} and references therein. The extraction of the symmetry energy from experiments \cite{Chen2004,tsang2012,PREX1,PREX2} is limited to quite symmetric nuclear matter. Astrophysical observations do not set strong enough constraints \cite{Oertel2017,Li2021}.  From the recent measurement of the neutron radius of 208Pb through parity-violation in electron scattering, PREX2 \cite{PREX2}, authors of \cite{Reed2021} obtained for the slope of the symmetry energy at saturation $L=106\pm37$ MeV, using the correlation between the neutron skin thickness and  the neutron EOS, which is directly related with the slope of the symmetry energy \cite{Brown2000}. The same data, together with  constraints determined from astrophysical  observations and microscopic calculations, were used to constraint the EOS of nuclear matter within a non-parametric approach \cite{Essick2021a,Essick2021b}. 

The knowledge of the density dependence of the symmetry energy is essential to determine the internal composition of a neutron star, in particular, the possible nucleation of the different types of hyperons \cite{Providencia2012,Providencia2018,Fortin2020} or the occurrence of a first order hadron-quark phase transition \cite{Ferreira:2020evu}. A direct effect of the symmetry energy can be observed on the cooling of NS, in particular, on the possible opening of direct Urca processes which give rise to a very fast NS cooling  \cite{Horowitz2002}. The nucleation of hyperons inside NS may explain a faster cooling inside NS than the prediction obtained from a nucleonic EOS \cite{Cavagnoli2011,Fortin2021}. 

In the present study, we will analyse the uncertainties associated with the extraction of nuclear matter properties as, for instance, properties related with the symmetry energy, from the knowledge of the $\beta$-equilibrium EOS. It will be shown that one of the main difficulties is precisely connected with the composition of neutron star matter. We will only analyse matter constituted by protons, neutrons, electrons and muons. 

In Section \ref{sec2}, we present the formalism and deduce the equation that determines the proton fraction as a function of the nucleonic density given the EOS of $\beta$-equilibrium matter. It is discussed the effect of (i)  the onset of muons, (ii) the approximation introduced when only  quadratic terms on the asymmetry parameter are kept in the EOS and (iii) taking a Taylor expansion around saturation density for the symmetric nuclear matter energy per particle and the symmetry energy. In Section \ref{sec3} we present some results, in particular, an estimation of the uncertainties introduced by each approximation. It is also discussed the possibility of expressing an EOS in terms of the Taylor expansion if the coefficients of the different terms are considered as effective parameters. Finally, in Section \ref{sec4} some conclusions are drawn.

\section{Formalism\label{sec2}}

In this section we will relate the $\beta$-equilibrium EOS with the nuclear matter EOS. The $\beta$-equilibrium DD2 EOS, an EOS obtained within a relativistic mean-field description with density dependent nucleon-meson couplings, will be considered as the reference \cite{Typel1999,Typel2009}.  In a first approach we will not include muons and will estimate the uncertainty introduced with this approximation. 

We  start from a general description, frequently applied, that considers a quadratic  dependence of the energy per particle of asymmetric nuclear matter on the asymmetry parameter $\delta =(\rho _{n}-\rho _{p})/(\rho _{n}+\rho _{p})$, where $\rho_n$ and $\rho_p$ are, respectively, the neutron and the proton densities \cite{Bombaci1991,Vidana2009}. Although the quadratic term on the asymmetry parameter $\delta$ was estimated in \cite{Vidana2009} to be quite small, we will also estimate the uncertainty introduced when it is neglected. In \cite{Xu2009} the authors have shown that the parabolic approximation may introduce huge errors in the determination of the NS crust-core transition density.

\subsection{Symmetry energy}

The energy per particle of asymmetric nuclear matter may be written in terms of the asymmetric parameter $\delta$ as
\begin{equation}{
\frac{E_{nuc}}{A}\left ( \rho,\delta  \right )= \frac{E_{SNM}}{A}\left ( \rho \right )+S\left ( \rho \right )\delta ^{2}+S_{4}\left ( \rho \right )\delta ^{4} + {\cal O}(\delta^6),} \label{EOS}\end{equation}
where $\rho=\rho_n+\rho_p$ is the baryonic density,  ${E_{SNM}}/{A}$ is the symmetric nuclear matter energy per particle, $S\left( \rho \right)$ is the symmetry energy
\begin{equation}
S\left ( \rho \right )=\frac{1}{2} \left.\frac{\partial^2 {E_{nuc}}/A}{\partial \delta^2}\right|_{\delta=0}
\end{equation}
and 
\begin{equation}
S_{4}\left ( \rho \right )= \frac{1}{4!} \left.\frac{\partial^4{E_{nuc}/A}}{\partial \delta^4}\right|_{\delta=0}.
\end{equation}

As in \cite{Vidana2009}, we expand in a Taylor expansion around the saturation density $\rho_0$ the symmetric nuclear matter energy per particle  ${E_{SNM}}/{A}$ and the symmetry energy $S(\rho)$. For the ${E_{SNM}}/{A}$, we obtain
\begin{equation}\frac{E_{SNM}}{A}\left ( \rho \right )=E_{0}+\frac{K_{0}}{2}\eta^{2}+\frac{J_{0}}{3!}\eta^{3}+\frac{Z_{0}}{4!}\eta^{4}+..., 
\label{snm}\end{equation}
with $\eta=({\rho-\rho_{0}})/({3\rho_{0}}),$ and
where the coefficient of the term of order $n$ is given by
\begin{equation}
\frac{X^{(n)}}{n!}\eta^n, \quad\mbox{with}\quad X^{(n)}=\left (3\rho _{0}  \right )^{n}\frac{d^{n} \left (E_{SNM}/A  \right )}{d \rho ^{n}}\bigg\rvert_{\rho =\rho _{0}}.
\label{coef}
\end{equation}
The coefficients of the four terms of lowest order  in Eq. (\ref{snm}) correspond to,
respectively, the energy per particle $E_0$, the incompressibility $K_0$, the skewness $J_0$ and the kurtosis $Z_{0}$ at saturation density. 
The expansion of the symmetry energy $S(\rho)$ around the saturation density is given by
\begin{equation}S\left ( \rho \right )=E_{sym}+L_{sym}\eta+\frac{K_{sym}}{2}\eta^{2}+\frac{J_{sym}}{3!}\eta^{3}+\frac{Z_{sym}}{4!}\eta^{4}+...
\label{esym}\end{equation}
where the coefficients of the expansion 
are calculated as in Eq. (\ref{coef}) replacing $E_{SNM}/A$ by $S$, and are identified, respectively, as  the symmetry energy  $E_{sym}$, and its slope $L_{sym}$, curvature $K_{sym}$, skewness  $J_{sym}$  and kurtosis $Z_{sym}$ at saturation density. 

Just for reference,  the values of the expansion coefficients associated with the model DD2, which has saturation density $\rho_0=0.149$ fm$^{-3}$, are given in Table \ref{tab:par}.

In the following, we show how the  asymmetry parameter $\delta$, or the  proton fraction $x_p=\rho_p/\rho$ since $\delta=1-2x_p$, can be calculated from  the symmetric nuclear matter energy per particle and the $\beta$- EOS, see Eq. (\ref{final}). The symmetry energy $S(\rho)$ will next be determined once $\delta$ is known. 

Let us consider the energy density of $\beta$-equilibrium matter $\varepsilon_\beta$, which we will take as the one obtained from the DD2 model, but which could as well be obtained from observations. Using a known EOS will allow us to  calculate  the uncertainties associated to the procedure. The relation of the energy density $\varepsilon_\beta$ with the nuclear matter energy per particle is given by
\begin{equation}
\varepsilon_\beta=\rho \left(\frac{E_{nuc}}{A}(\rho,\delta)+\bar{m}_N\right)+\varepsilon_{lep}(\rho,\delta), 
\label{ebeta}
\end{equation}
where $\varepsilon_{lep}$ designates the leptonic energy density, which for cold-catalysed matter includes the contribution of electrons and muons
\begin{equation}
\varepsilon_{lep}(\rho,\delta)= \varepsilon_{e}(\rho,\delta)+\varepsilon_{\mu}(\rho,\delta),
\end{equation}
and  
$\bar{m}_N=m_n(1-x)+m_p x$, with $x=\rho_p/\rho$ the proton fraction,  which reduces to the nucleon vacuum mass $m_N$ if   $m_n$ and $m_p$, respectively, the neutron and the proton masses are taken equal to the average nucleon mass $(m_n+m_p)/2$, which we designate by $m_N$ in the following. This approximation is generally considered in relativistic mean field (RMF) models.

Charge neutrality establishes a relation between $\rho_p$, $\rho_e$ and $\rho_\mu$. If no muons are present
\begin{equation} \rho_p=\rho_e, \quad \mbox{or}\quad k_{Fp}=k_{Fe}, \label{charge1}\end{equation}
where $k_{Fi}$ is the Fermi momentum of particle $i$. After the onset of muons the above relations are replaced by 
\begin{equation} \rho_p=\rho_e+\rho_\mu \quad \mbox{or}\quad k^3_{Fp}=k^3_{Fe}+k^3_{F\mu}. \label{charge2}\end{equation}
The energy density of a gas of free non-interacting relativistic electrons or muons is
\begin{equation}\varepsilon _{i}=\frac{m_{i}^{4}}{8\pi ^{2}}\left ( y_{i}\left ( 2y_{i}^{2}+1 \right )\sqrt{y_{i}^{2}+1} -ln\left ( y_{i} +\sqrt{y_{i}^{2}+1}\right )\right ),
\label{EOSlep}\end{equation}
$i=e,\, \mu,$ with $y_{i}={k_{Fi}}/{m_{i}}$.

$\beta$-equilibrium imposes the following relations between the chemical potentials of neutrons, protons, electrons and muons
\begin{equation}\mu _{n}=\mu _{p}+\mu _{e}, \qquad \mu_\mu=\mu_e,\end{equation}
with 
\begin{equation}\mu_{p}=\frac{\partial \varepsilon _{nuc}}{\partial \rho _{p}}, \quad \mu_{n}=\frac{\partial \varepsilon _{nuc}}{\partial \rho _{n}},\end{equation}
and
\begin{equation}\varepsilon_{nuc} =\rho \left ( \frac{E_{nuc}}{A}+m_n(1-x)+m_p x \right ).\end{equation}
Performing the derivatives, the following relation is obtained 
\begin{equation}\mu _{n}-\mu _{p}=2\frac{\partial\left ( E_{nuc}/A \right ) }{\partial \delta }+m_n-m_p=\mu_e.
\label{cheq}
\end{equation}
or, if  $S_4$ in Eq. (\ref{EOS}) is put to zero
\begin{equation}\frac{\partial \left ( E_{nuc}/A \right )}{\partial \delta }= 2\delta\,  S\left (\rho  \right ),\end{equation}
so
\begin{equation}\mu _{n}-\mu _{p}= \mu_e=4\delta\,  S\left (\rho  \right )+m_n-m_p.
\label{cheq1}
\end{equation}
This equation will be used later to determine $S(\rho)$ from the knowledge of the electron chemical potential $\mu_e$ and the asymmetry parameter $\delta$.

Taking Eq. (\ref{ebeta}), we express ${E_{nuc}}/{A}$ in terms of the $\beta$-equilibrium EOS, $\varepsilon_\beta$, and the energy density of the leptons
\begin{equation}\frac{E_{nuc}}{A}\left ( \rho ,\delta  \right )=\frac{\varepsilon _{\beta }\left ( \rho  \right )-\varepsilon _{e}\left ( \rho ,\delta  \right )-\varepsilon _{\mu }\left ( \rho ,\delta  \right )}{\rho }-m_{N}.
\label{EOS1}\end{equation}
Considering the nuclear energy written as in Eq. (\ref{EOS}), the symmetry energy $S(\rho)$ is expressed in terms of the total nuclear matter energy per particle and the symmetric nuclear matter energy per particle
\begin{equation}
 S\left (\rho  \right )= \frac{1}{\delta^2 }\left (  \frac{E_{nuc}}{A}\left ( \rho ,\delta  \right )-\frac{E_{SNM}}{A}\left ( \rho  \right )\right ).\label{EOS2}
\end{equation}
From Eqs. (\ref{EOS1}), (\ref{EOS2}) and (\ref{cheq1}) we obtain the relation (see \cite{Essick2021}),
\begin{eqnarray}
\frac{\delta}{4}\left( m_p-m_n+\mu_e\right)&=& \nonumber\\
\frac{\varepsilon _{\beta }\left ( \rho  \right )-\varepsilon _{e}\left ( \rho ,\delta  \right )-\varepsilon _{\mu }\left ( \rho ,\delta  \right )}{\rho }-m_{N}-\frac{E_{SNM}}{A},
\label{final}
\end{eqnarray}
which includes the contribution of muons. This equation should be supplemented  by Eq. (\ref{EOSlep}) together with the charge neutrality conditions, Eq. (\ref{charge1}) or (\ref{charge2}), and the expression of the electron and muon chemical potentials 
\begin{equation}\mu_i=\sqrt{k^2_{Fi}+ m^2_i}, \quad i=e,\mu
\label{me}
\end{equation}
with $m_e\, (m_\mu)$ the mass of the electron (muon).

Eq. (\ref{final}) is the relation we need to calculate the asymmetry parameter $\delta$, and from Eq. (\ref{cheq1}) obtain the symmetry energy $S(\rho)$.

The inclusion of the fourth order term on the asymmetry parameter $\delta$ in Eq. (\ref{EOS})  introduces an extra term $-S_4(\rho) \delta^4$ on the right side of Eq. (\ref{final}). If a RMF model with no terms of an higher order than quadratic in the $\rho$-meson is used (this is the case of DD2),  the only contribution to the $S_4$ term  comes from the kinetic terms.   In this case
\begin{equation}S_{4}\left ( \rho  \right )=\frac{1}{4! 3^3}
\frac{10\,k_{F}^{6}+11\,k_{F}^{4}m^{*2}+4\,k_{F}^{2}m^{*4}}{\epsilon _{F}^{5}},
\label{s4}
\end{equation}
where $\epsilon_F=\sqrt{k_F^2+m^{*2}}$, $k_F$ is the Fermi momentum of symmetric nuclear matter associated with the density $\rho$, $k_F=\left(3\pi^2 \rho/2\right)^{1/3}$, {and $m^*$ is the Dirac effective mass.}
In the following we will also show the contribution of the sixth  and eighth order terms in $\delta$, (see Eqs. (\ref{s6}) and (\ref{s8}) in the Appendix.\\

\section{Results \label{sec3}}
{In this section, we present and discuss the uncertainties introduced by several approximations when extracting information from a given NS EOS in $\beta-$equilibrium.}

\subsection{Testing Eq. (\ref{final})}

In the following, we discuss several approximations to Eq. (\ref{final}), and also the approximation in which it is based, i.e., the possibility of writing the EOS keeping only terms independent  of  $\delta$  and quadratic in $\delta$, Eq. (\ref{EOS}).  This equation has been used before to extract the proton fraction from a $\beta$-equilibrium EOS, see, for instance \cite{Essick2021}, and to build the $\beta$-equilibrium EOS from the knowledge of the neutron matter EOS and symmetric nuclear matter EOS, see \cite{Bombaci1991,Hebeler2013}. We assume that the difference between the proton and neutron masses is small compared with the electron chemical potential, and the same is assumed when calculating the exact DD2 $\beta$-equilibrium EOS, $\varepsilon_\beta$, that enters in Eq. (\ref{final}).
 Considering these assumptions, we reach the following relation
\begin{equation}
\frac{\delta}{4}\mu_e=\frac{\varepsilon _{\beta }^{e,\mu}\left ( \rho  \right )-\varepsilon _{e}\left ( \rho ,\delta  \right )-\varepsilon _{\mu}\left ( \rho ,\delta  \right )}{\rho }-m_{N}-\frac{E_{SNM}}{A}.
\label{final1}
\end{equation}
To simplify our following discussion, we have denoted the  DD2 energy density of matter in $\beta-$equilibrium by $\varepsilon\equiv \varepsilon _{\beta }^{e,\mu}$, where the superscript indices indicate which leptons were considered, $\varepsilon _{\beta }^{e}$ if just electrons are considered and $\varepsilon _{\beta }^{e,\mu}$ if both electrons and muons are included. This allows us to easily distinguish between the leptons considered in the calculation of the  DD2 energy density, $\varepsilon _{\beta }^{e,\mu}$, and the leptonic contributions entering Eq. (\ref{final1}), i.e.,  $\varepsilon _{e}$ and $\varepsilon _{\mu}$.
Let us first analyze the case without muons, i.e., $\varepsilon_{\mu}=0$.
\begin{figure}[!t]
	\includegraphics[width=0.95\linewidth]{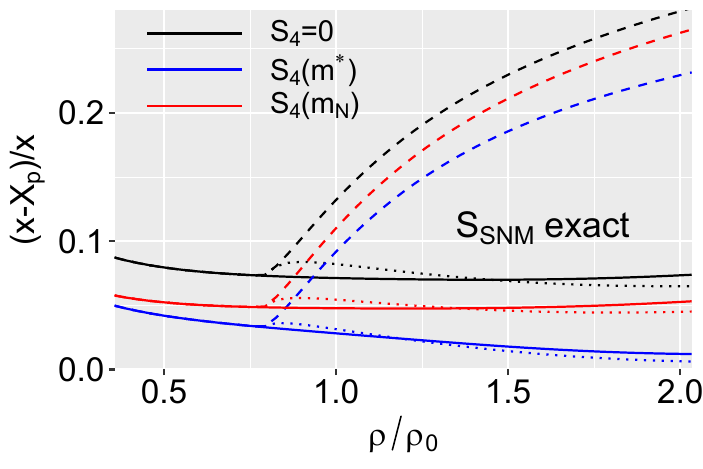}
	\caption{Relative deviation between the proton fraction $x_p$ calculated using Eq. (\ref{final1}) and the exact one $x$ given by the DD2 model, considering in Eq. (\ref{final1}): no contribution from $S_4$ (black), $S_4$ calculated with the effective mass $m^*$ (blue), and $S_4$ calculated with the vacuum mass $m_N$ (red). The line-type indicates which leptons are considered in the Eq. (\ref{final1}), when extracting $x_p$, and in the reference model (DD2) to calculate $x$: $\varepsilon_{\mu}=0$ and  $\varepsilon _{\beta }^{e}$ (solid), $\varepsilon_{\mu}=0$ and $\varepsilon _{\beta }^{e,\mu}$ (dashed), $\varepsilon_{\mu}\neq0$ and $\varepsilon _{\beta }^{e,\mu}$ (dotted).
}
	\label{fig1}
\end{figure}
In Fig. \ref{fig1}  we plot the relative deviation of the exact proton fraction  $x$ for cold matter in $\beta$-equilibrium described by the DD2 model and the proton  fraction $x_p$ obtained as a solution of EOS (\ref{final1}). In order to isolate the approximations arising solely from Eq. (\ref{final1}),  we use the exact value of the DD2 model for $E_{SNM}/A$.

 A comparison between $x$ and $x_p$ allows one  to discuss the errors introduced when Eq. (\ref{final1}) is used to calculate the exact proton fraction.
In our particular study, we take the  energy density  $\varepsilon _{\beta }$ obtained within DD2 model, but in a general case, $\varepsilon _{\beta }$  would be the  EOS  extracted from the measurement of the radius and mass of several NS.  It encloses information on the properties of hadronic matter and the different species of particles that nucleate inside the NS. For the densities considered, from just above the crust-core transition to two times the saturation density, the species that are expected to exist inside the NS are protons, neutrons and electrons, and above $\rho\approx$ 0.11-0.12 fm$^{-3}$, also muons. Full (dashed) lines where obtained considering that the exact DD2 $\beta$-equilibrium EOS  does not include  (includes) muons,  which we designate by  $\varepsilon _{\beta }^{e}$ ($\varepsilon _{\beta }^{e,\mu}$). The top black curves describe these two situations. At saturation density, the relative error introduced is $\approx 7\%$ (full line) if the  $\beta$-equilibrium EOS does not consider muons. If, however, the $\beta$-equilibrium DD2 EOS contains muons in the star ($\varepsilon _{\beta }=\varepsilon _{\beta }^{e,\mu}$), we would have obtained the dashed line from Eq. (\ref{final1}), corresponding to a deviation at saturation density of $\approx 14\%$,  the double of the previous error.

In order to understand why the uncertainty is as high as $\approx 7\%$ at saturation density, even using the exact EOS for symmetric nuclear matter, it should be recalled that terms of order  ${\cal O}(\delta^4)$ or higher have been neglected in Eq. (\ref{EOS}). In model DD2 the interaction is quadratic in the asymmetry parameter $\delta$, therefore,  a higher order contribution may only come from the kinetic term as discussed in the previous section. $S_4$ is given by Eq. (\ref{s4}) in this case. This expression includes an interaction dependent quantity, the effective mass $m^*$,  which in principle is not known. In order to estimate the effect of $S_4$, we solve Eq. (\ref{final1}) including the $S_4$ term
\begin{equation}
\frac{\delta}{4}\mu_e=\frac{\varepsilon _{\beta }^{e,\mu}\left ( \rho  \right )-\varepsilon _{e}\left ( \rho ,\delta  \right )-\varepsilon _{\mu}\left ( \rho ,\delta  \right )}{\rho }-m_{N}-\frac{E_{SNM}}{A}- S_4 \delta^4,
\label{final2}
\end{equation}
calculated with the effective mass $S_4(m^*)$ (blue) and with the nucleon vacuum mass $m_N$, $S_4(m_N)$ (red).  With the new term,  the error at saturation is reduced to 5\% using the vacuum mass, and to 3\% with the effective mass. 
The dotted curves in Fig. \ref{fig1} where obtained keeping
the muon term $\varepsilon _{\mu}$ in Eq. (\ref{final1}), and, as before, using  the exact DD2 model with muons, i.e., $\varepsilon _{\beta }^{e,\mu}$. It is interesting to note that, within a good approximation, the dotted lines almost coincide with the full lines which totally exclude the presence of muons both in the $\beta$-equilibrium EOS and 
in Eq. (\ref{final1}).\\

\begin{figure}[!t]
\begin{tabular}{c}
\includegraphics[width=0.95\linewidth]{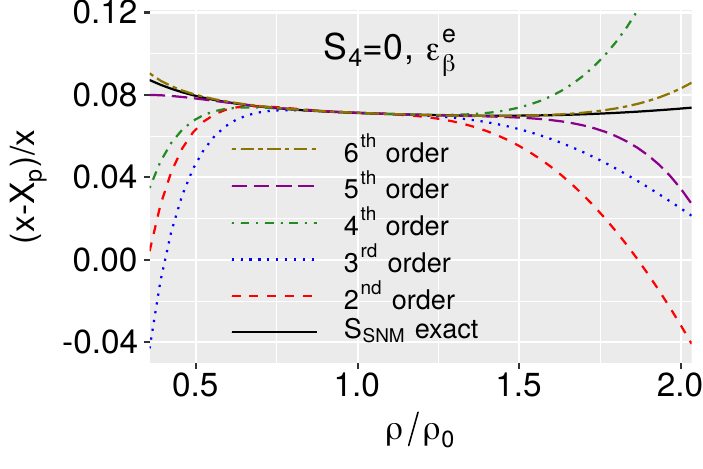}\\	
\includegraphics[width=0.95\linewidth]{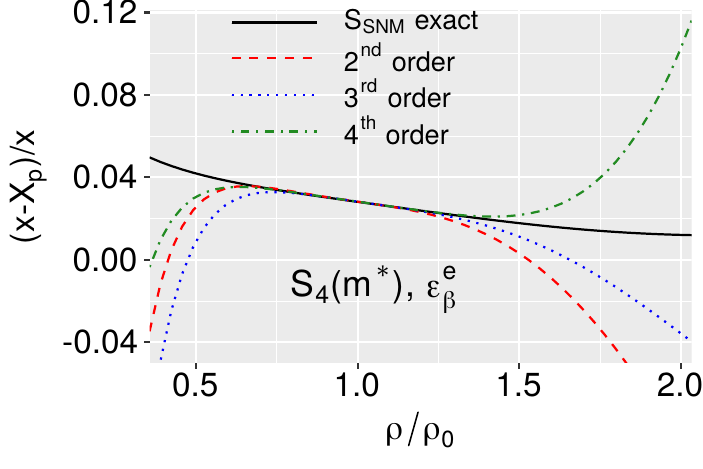}\\
\includegraphics[width=0.95\linewidth]{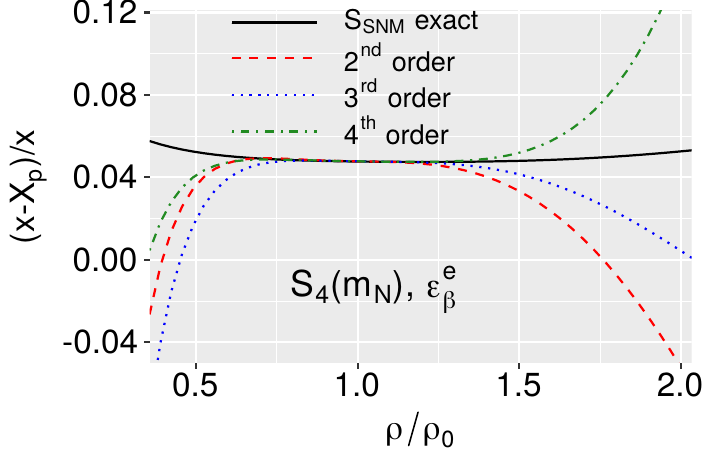} 
\end{tabular}
	\caption{Relative deviation between the proton fraction $x_p$  [Eqs. (\ref{final1})-(\ref{final2})] and the exact one $x$ given by the DD2 model, for the different approximations defined in Eq. (\ref{snm}).
	Muons were not considered,  i.e., $\varepsilon_{\mu}=0$ and $\varepsilon _{\beta }^{e}$.	Top panel: the $S_4$ term and higher orders were neglected, middle panel: the $S_4$ term was included using the vacuum mass, bottom panel: the the $S_4$ term was included using the saturation effective mass.
	}
	\label{fig2}
\end{figure}

Next, we consider an approximation for the symmetric matter EOS, ${E_{SNM}}/{A}$. Nuclear matter properties are quite well restricted at saturation density and, therefore, a Taylor expansion around saturation density will allow one to describe reasonably well the EOS in a range of densities close to saturation density. We solve Eqs. (\ref{final1}) and (\ref{final2}) taking the expansion given in Eq. (\ref{snm}) until second, third and fourth order. 
The results are shown in Fig. \ref{fig2}, where the top panel also includes curves obtained considering terms until fifth and sixth order.
It is seen that in the range of densities $0.8\rho/\rho_0-1.2 \rho/\rho_0$, all approximations coincide with the exact ${E_{SNM}}/{A}$ (top panel). Taking the fourth order, the error introduced is below 3\% between  $\approx0.4\rho/\rho_0-1.8 \rho/\rho_0$, and at the sixth order the error is below 1\% in the range $\approx0.3\rho/\rho_0-2 \rho/\rho_0$. With the second order approximation, an error of the order of 3\% is possible but within a smaller density interval, $\approx 0.5\rho/\rho_0-1.5 \rho/\rho_0$. Similar conclusions are drawn when the term $S_4$ is included, see Fig. \ref{fig2} middle and bottom panels.\\

\begin{figure}[!t]
\begin{tabular}{c}
\hspace*{-.5cm}	
\includegraphics[width=.95\linewidth]{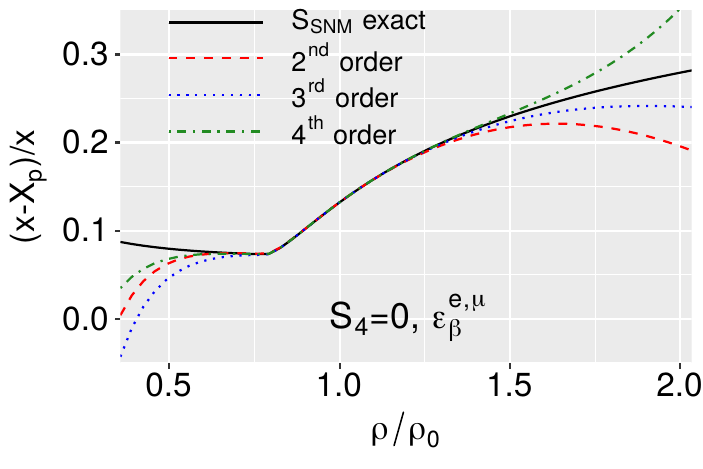}\\ 	
\includegraphics[width=.95\linewidth]{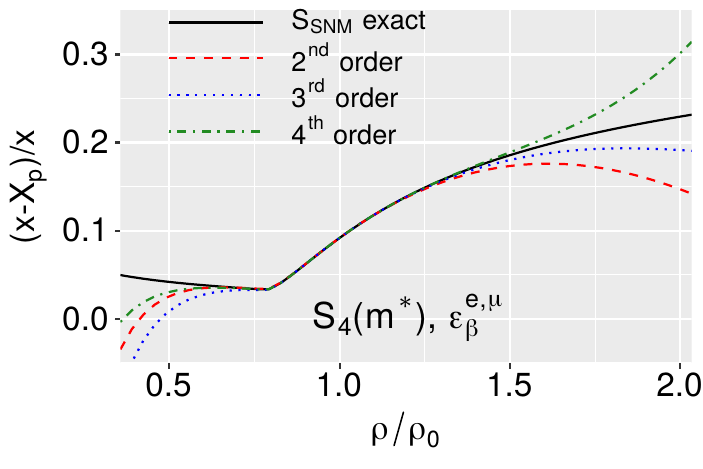} \\	
\includegraphics[width=.95\linewidth]{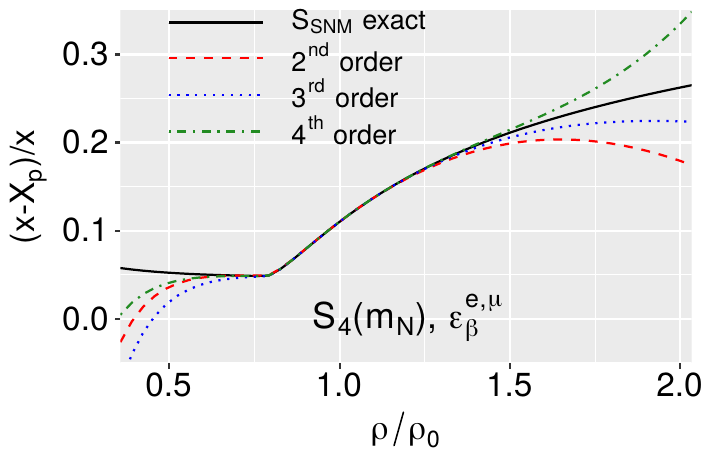} 
\end{tabular}
	\caption{The same as in Fig. \ref{fig2} but now 
	considering that muons were included ($\varepsilon _{\beta }^{e,\mu}$) when determining $x$  with DD2 model but not in the determination of $x_p$ ($\varepsilon_{\mu}=0$).
	See Eqs.  (\ref{final1})-(\ref{final2}). 
	} 
	\label{fig3}
\end{figure}

\begin{figure}[!t]
\begin{tabular}{c}
\hspace*{-.5cm}	
\includegraphics[width=.95\linewidth]{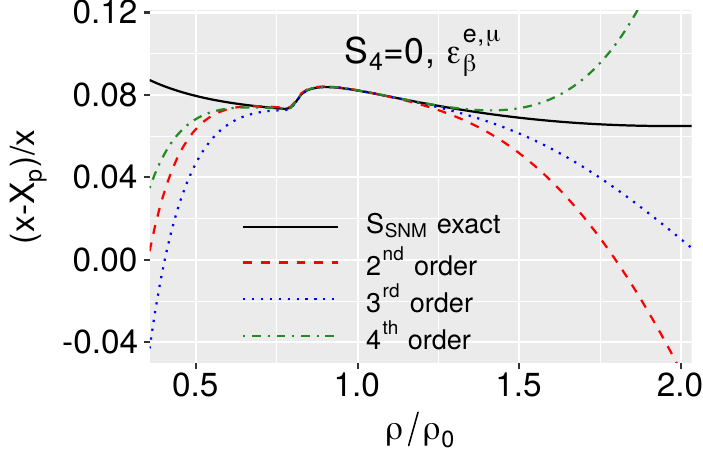} \\
\includegraphics[width=.95\linewidth]{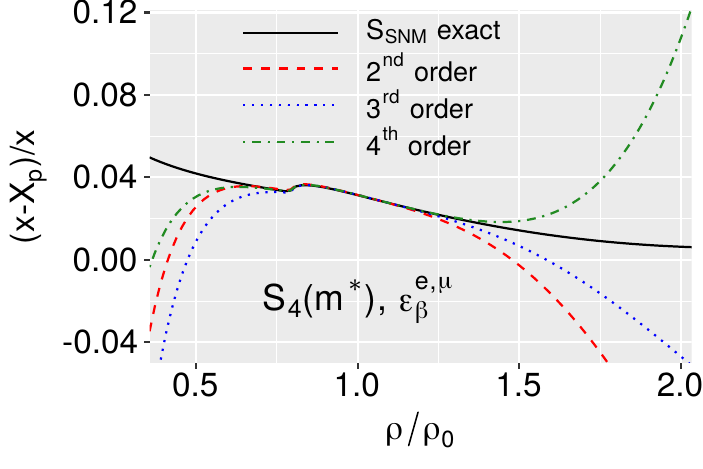} \\	
\includegraphics[width=.95\linewidth]{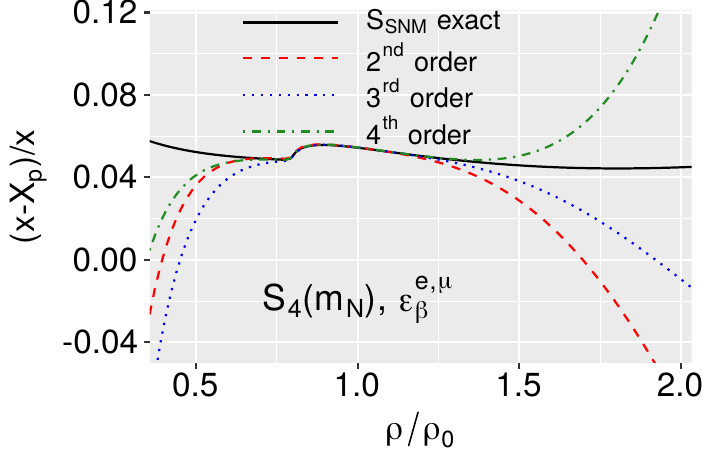} 
\end{tabular}
	\caption{The same as in Fig. \ref{fig2} but now 
	considering that muons were included  both when determining $x$ ($\varepsilon _{\beta }^{e,\mu}$) with DD2 model and in the determination of $x_p$ ($\varepsilon _{\mu}\neq0$).
	See Eqs.  (\ref{final1})-(\ref{final2}).}
	\label{fig4}
\end{figure}

In Fig. \ref{fig3}, the exact proton fraction $x$  has been calculated considering that the $\beta$-equilibrium DD2 EOS contains muons in the star ($\varepsilon _{\beta }=\varepsilon _{\beta }^{e,\mu}$), and considering  Eq. (\ref{final1}) to get the top panel, and Eq. (\ref{final2}) to obtain the middle and bottom panels. Once more, the Taylor expansion coincides with the exact solution for symmetric nuclear matter in a small range of densities close to saturation. However, as already discussed above, the error introduced by not considering explicitly the presence of muons in Eqs.  (\ref{final1})  and (\ref{final2}) increases the error in the estimation of the proton fraction to values above 10\% at saturation density and more than 20\% at 1.5$\rho_0$.  The inclusion of the term $S_4(m^*)$  reduces the error to about 5\% at saturation.  Repeating the same exercise, but this time keeping the term $\varepsilon_\mu$ in Eq. (\ref{final1}), we have obtained the results plotted in Fig. \ref{fig4}. The uncertainties reduce to values similar to the ones shown in Fig. \ref{fig2} when muons are not considered in Eq. (\ref{final1}) and in the determination of the $\beta$-equilibrium EOS, $\varepsilon _{\beta }^{e}$.

\subsection{Extracting the symmetry energy}

One of the main interests on NS is the possibility of obtaining information on very asymmetric nuclear matter.  Once the proton fraction is known, Eq. (\ref{cheq1}) allows one to determine the symmetry energy $S(\rho)$. This exercise was done within the different approximations discussed in the previous section.

\begin{figure}[!t]
\begin{tabular}{c}		
\includegraphics[width=.95\linewidth]{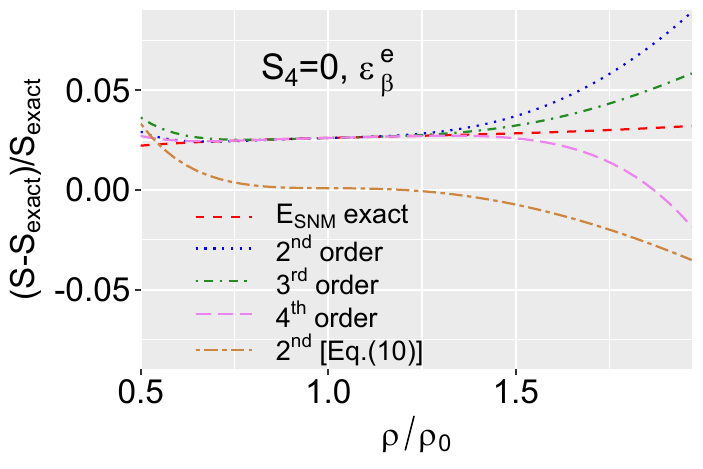} \\	
\includegraphics[width=.95\linewidth]{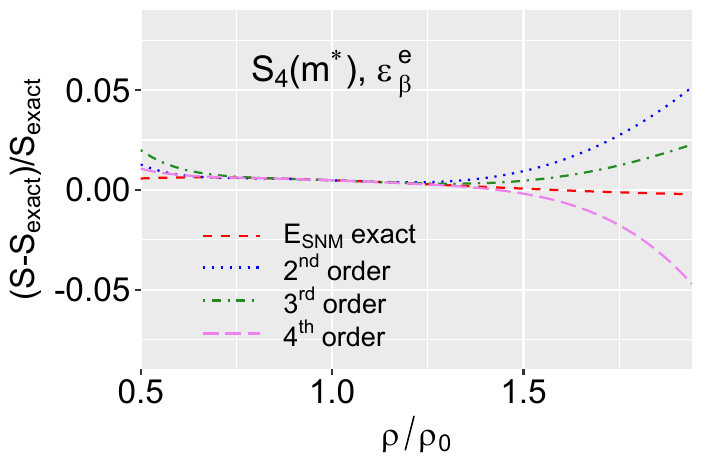} \\	
\includegraphics[width=.95\linewidth]{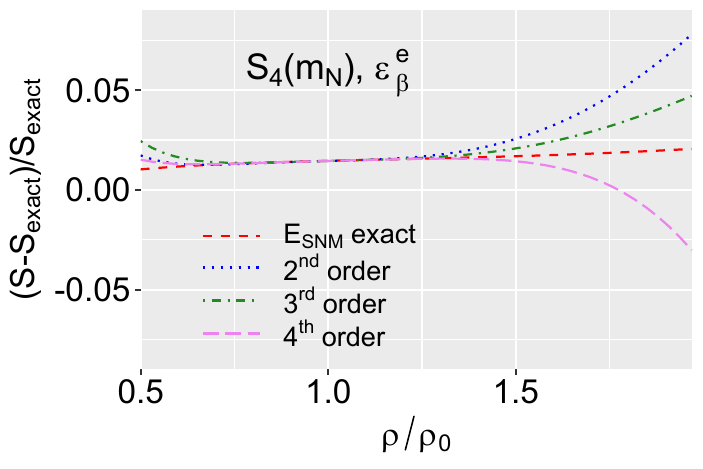} 
\end{tabular}
	\caption{The exact DD2 symmetry energy (black line) is compared with the symmetry energy calculated from Eq. (\ref{final1}) with $\varepsilon_\mu=0$ and with different approximations for the symmetric nuclear matter EOS, $E_{SNM}/A$ (see text for details). The brown line in the top panel was obtained taking the DD2 expansion terms up to second order in Eq. (\ref{esym}).}
		\label{fig:esym}
\end{figure}

 In Fig. \ref{fig:esym}, the
 relative deviation of the symmetry energy  from the exact DD2 symmetry energy  with the symmetry energy determined from: (i) Eq. (\ref{esym}) taking the DD2 model terms up to second order (brown line, shown only in the top panel); (ii) Eq. (\ref{cheq1}) once the proton fraction is known [by solving Eqs. (\ref{final2}) or (\ref{final1}), depending on whether the term $S_4(\rho)$ is considered or not] assuming for $E_{SNM}/A$ the exact value or the successive order approximations, given by Eq. (\ref{snm}). These results were obtained not considering the muons in the DD2 model, i.e., with $\varepsilon _{\beta }^{e}$.
 
Three scenarios for the fourth order terms on the asymmetry parameter are
analyzed: $S_4(\rho)=0$ (top panel), $S_4(\rho)$ calculated with the exact effective mass (middle) or with the vacuum nucleon mass (bottom). Neglecting the fourth order term $S_4$ gives rise to a deviation from the exact result of the order of +3\%. This increase of the symmetry energy is clear from Eq. (\ref{EOS2}): neglecting $S_4(\rho)$ and other higher terms gives a larger $S(\rho)$, since these terms are all positive and they should all subtract from the energy per particle $E_{nucl}/A$. This difference is reduced to 1\% when $S_4(\rho)$ calculated with the vacuum mass is considered. Taking for the symmetric nuclear matter EOS  terms until second order in $\eta$ in Eq. (\ref{snm}) is enough to give deviations below 3\% for densities in the range $0.5\rho_0-1.8\rho_0$.\\

\begin{figure}[!t]
\begin{tabular}{c}
\includegraphics[width=0.95\linewidth]{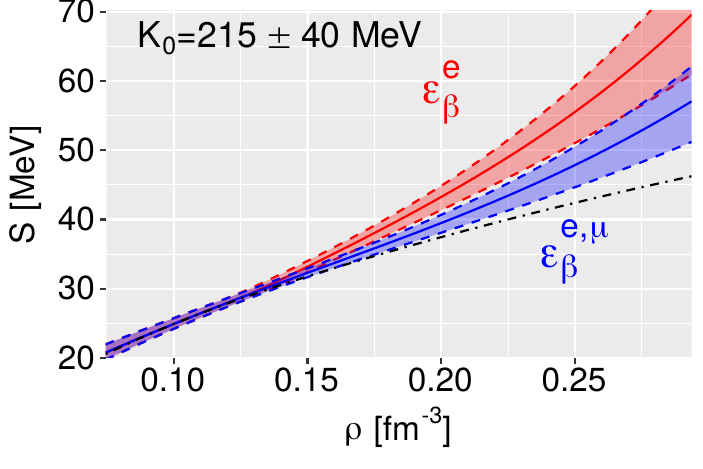}\\
\includegraphics[width=0.95\linewidth]{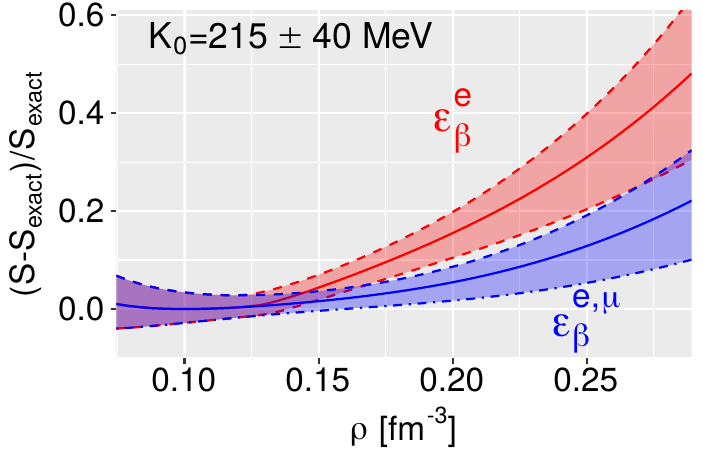} 
\end{tabular}
	\caption{Symmetry energy (top panel) and relative deviation from the exact symmetry energy (bottom panel) predicted by Eqs. (\ref{cheq1}) and (\ref{final1}) with $\varepsilon_\mu=0$ and using for $E_{SNM}$ a Taylor expansion up to second order with $\rho_0=0.164$ fm$^{-3}$, $E_{0}=15.86$ MeV  and $K_{0}=215$ MeV (middle line) and $\rho_0=0.171 (0.158)$ fm$^{-3}$, $E_{0}=16.43 (15.29)$ MeV and $K_{0}=175(255)$ MeV define the bottom (top) limits \cite{Huth2020}. The blue (red) bands considers $\varepsilon _{\beta }^{e,\mu}$ ($\varepsilon _{\beta }^{e}$). 
}
	\label{fig:esym2}
\end{figure}

However,  the symmetric nuclear matter EOS is only known within a given uncertainty. Taking for the saturation density, the binding energy  and the incompressibility at saturation, the values obtained from a $\chi$EFT calculation \cite{Drischler2017},  respectively, $\rho_0=0.165\pm 0.007$ fm$^{-3}$, $E_{0}=15.86\pm 0.57$ MeV
and $K_{0}=215\pm40$ MeV \cite{Drischler2017}, we have obtained the symmetry energy as a function of the density, see Fig. \ref{fig:esym2}.  Eqs. (\ref{cheq1})  and (\ref{final1}) were used with $\varepsilon_\mu=0$, using for $E_{SNM}/A$ a Taylor expansion up to second order. In the top panel of Fig. \ref{fig:esym2}, the predicted symmetry energy is compared with the DD2 symmetry energy \mf{(dot-dashed line)}. To quantify the differences,   we  show the relative differences of the deduced symmetry energy with respect to the exact one in the bottom panel.
The scenario with muons inside the NS, i.e.  $\varepsilon _{\beta }=\varepsilon _{\beta }^{e,\mu}$, is displayed in blue, while the red curve represents the absence of muons in the calculation of the NS EOS with the DD2 model, $\varepsilon _{\beta }=\varepsilon _{\beta }^{e}$.
Since the presence of muons decreases the fraction of electrons, it is clear that the calculation with muons predicts a smaller symmetry energy, and therefore, larger proton-neutron asymmetries  are attained. It is seen that the predicted symmetry energy from  Eqs. (\ref{cheq1})  and (\ref{final1}) overshoots the exact symmetry energy above saturation density. This occurs because above the saturation density  the predicted proton fraction $x_p$ is larger than the exact one $x$ if  the Taylor expansion is cut at second order, see Fig. \ref{fig:esym}.   A larger proton fraction is associated with a larger symmetry energy, as observed. Note that the separation between  both bands occurs at the muon onset.

\subsection{Taylor expansion as a functional}

It was shown that  Eq. (\ref{EOS}), with  $E_{SNM}/A$ and $S(\rho)$ given by the Taylor expansions  defined in Eqs. (\ref{snm}) and  (\ref{esym}),   is adequate to reproduce the nuclear matter EOS only in a small range of densities not far from the saturation density $\rho_0$ and below 2$\rho_0$.
However, in \cite{Margueron2017} the authors demonstrate that a realistic EOS can be  parametrized to a good approximation by a fourth order Taylor expansion of both the symmetric nuclear matter EOS and symmetry energy. To achieve this, it is necessary  to know the energy density and respective proton fraction at a given high density point. This is interesting, because it indicates that a smooth EOS, as one that describes nucleonic nuclear matter, can be parametrized by a Taylor expansion. It is, however, important to point out that the higher order parameters are effective and do not have the meaning one gives to the expansion coefficients associated with the nuclear matter properties at saturation. This type of parametrization for the nuclear matter EOS has been used in several recent studies \cite{Margueron2017,Margueron:2017lup,Zhang2018,Xie2019,Ferreira:2019bgy,Ferreira:2021pni}.\\

\begin{figure}[!t]
\begin{tabular}{c}
\includegraphics[width=0.95\linewidth]{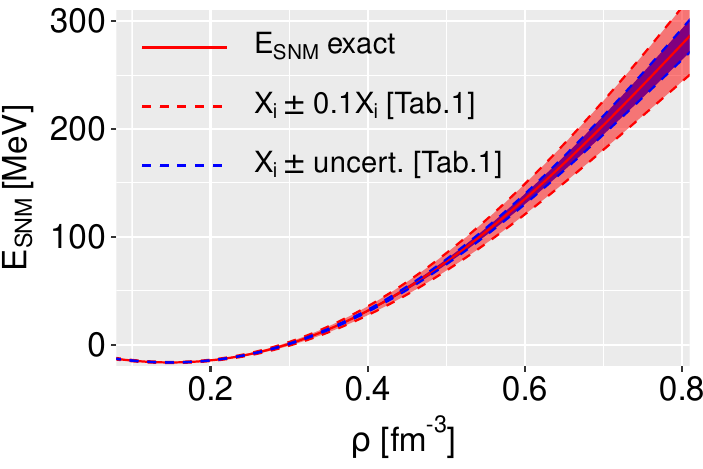}\\ 
\includegraphics[width=0.95\linewidth]{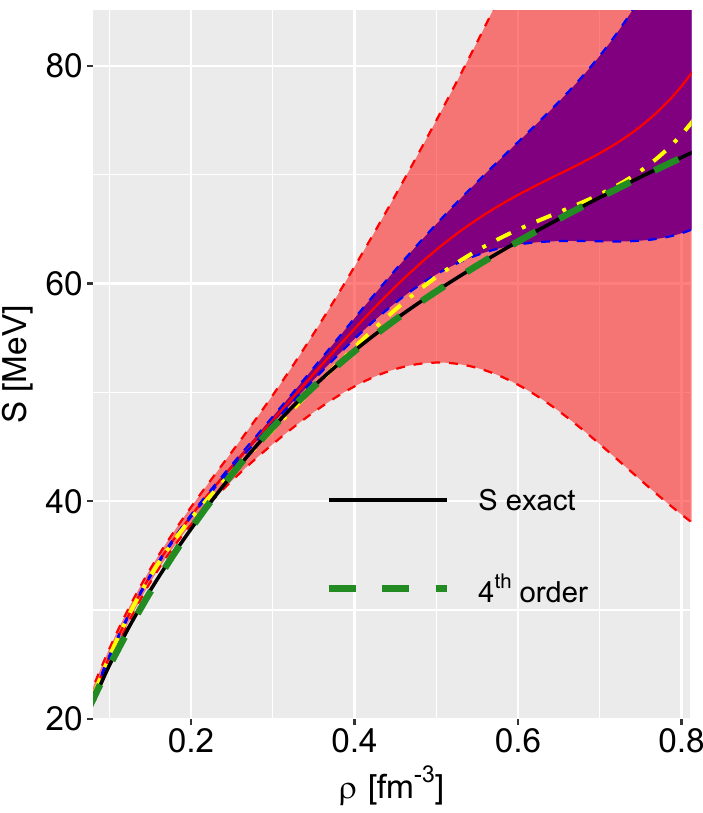}\\ 
\end{tabular}
	\caption{Top panel: Symmetric nuclear matter energy per particle  $E_{SNM}/A$; bottom panel:  symmetry energy  extracted from the Eq. (\ref{final1}) with $\varepsilon_\mu=0$. For $E_{SNM}/A$ a Taylor expansion functional with terms until fourth order fitted to the DD2 $E_{SNM}/A$ in the range of densities $0.08<\rho<0.8$ fm$^{-3}$ was considered,  see Table \ref{tab:par}. The dashed green line represents a fourth order expansion for $S(\rho)$ [Eq. (\ref{esym})] with the parameters defined in Table \ref{tab:par}.
	The yellow dashed-dotted and solid red lines show $S(\rho)$ when muons are or are not considered ($\varepsilon _{\beta }^{e,\mu}$ or $\varepsilon _{\beta }^{e}$), respectively, in Eq. (\ref{final1}). 
	}
	\label{fig:S3}
\end{figure}

In the following, we suppose that the expansions defined by Eqs.  (\ref{snm}) and (\ref{esym}) can be considered  adequate functionals to parametrize a given EOS. We have taken the symmetric nuclear matter EOS, $E_{SNM}/A$, until fourth and sixth order given by
\begin{eqnarray}
\frac{{E'}_{SNM}}{A}\left ( \rho \right)
&=&{E'}_{0}
+\frac{{K'}_0}{2}\eta^{2}
+\frac{{J'}_{0}}{3!}\eta^{3}
+\frac{{Z'}_{0}}{4!}\eta^{4}\nonumber\\
&+&
\frac{{X^{(5)'}}_{0}}{5!}\eta^{5}
+\frac{{X^{(6)'}}_{0}}{6!}\eta^{6}
\label{snm1}
\end{eqnarray}
and the symmetry energy until fourth order 
\begin{equation}S'\left ( \rho \right )=E'_{sym}+L'_{sym}\eta+\frac{K'_{sym}}{2}\eta^{2}+\frac{J'_{sym}}{3!}\eta^{3}+\frac{Z'_{sym}}{4!}\eta^{4}.
\label{esym1}\end{equation}
We have then fitted the parameters of the expansions to the exact DD2 $E_{SNM}/A$ and $S(\rho)$ functions in two density ranges: $0.07<\rho<0.8$ fm$^{-3}$ and $0.07<\rho<0.32$ fm$^{-3}$.  In Table \ref{tab:par} the fitted parameters are given together with the associated uncertainties. Also shown are the  DD2 parameters. Some conclusions are in order: (i) taking enough terms in the expansion, in particular, the sixth order in $E_{SNM}/A$ and the fourth order in $S$, and performing the fit in the density range $0.07<\rho<0.8$ fm$^{-3}$, the lower order coefficients ${E'}_{0}$, ${K'}_0 $, ${E'}_{sym}$ and ${L'}_{sym}$ are very close to the DD2 parameters; (ii) if the fit is restricted  to a smaller density range,  $0.07<\rho<0.32$ fm$^{-3}$, also $J'_0$ and $K'_{sym}$ come out very close to the respective DD2 parameters. These results seem to indicate that 
taking enough terms in the expansion of the symmetric nuclear matter EOS and the symmetry energy, the lower order coefficients are very close to the nuclear matter properties at saturation. However, 
the coefficients of the higher order terms are effective and take into account the missing higher order terms. They should not be considered representative of the respective nuclear matter saturation properties. It is clear that if the density range used to perform the fit is smaller, with an upper limit closer to the saturation density, the higher order coefficients are closer to the exact ones.

In Fig. \ref{fig:S3} bottom panel, the exact DD2 symmetry energy  (black full line) is compared with the symmetry energy  obtained taking for $E_{SNM}/A$ the function represented in Fig. \ref{fig:S3} top panel, i.e. a Taylor expansion functional with terms until fourth order. The parameters used are given in Table \ref{tab:par}  and correspond to the ones obtained in the fit taking $\rho=0.8$ fm$^{-3}$ as the upper limit.  Eq. (\ref{final1}) was solved with $\varepsilon_\mu=0$ to obtain the proton fraction $x_p$ and Eq. (\ref{cheq1}) was next applied to determine the symmetry energy from the proton fraction. For the $\beta$-equilibrium EOS we have considered a calculation performed without muons (full red line), $\varepsilon _{\beta }=\varepsilon _{\beta }^{e}$, and with muons (yellow dashed-dotted line), $\varepsilon _{\beta }=\varepsilon _{\beta }^{e,\mu}$.
We also show the symmetry energy fit using a fourth order expansion with the parameters defined in Table \ref{tab:par} considering the  upper density limit 0.8 fm$^{-3}$ (green dashed line). It is striking that it coincides with the exact one. It is interesting that both  previsions obtained  from Eq. (\ref{final1}) (with $\varepsilon _{\beta }^{e}$ and $\varepsilon _{\beta }^{e,\mu}$) do not deviate much from the exact curve. This seems to indicate that the isoscalar and isovector channels are quite independent so that giving a reasonable $E_{SNM}/A$ EOS allows one to extract the symmetry energy from $\varepsilon _{\beta }$.

Including the muon in the $\beta$-equilibrium EOS, $\varepsilon _{\beta }=\varepsilon _{\beta }^{e,\mu}$,   reduces the proton fraction, and, therefore, after the muon onset the symmetry energy decreases relative to the one obtained in a calculation without muons. The bands included in the figure represent the spread generated when the uncertainties on the parameters of $E_{SNM}/A$ given in Table \ref{tab:par} are taken into account (purple band).  We have also considered a larger uncertainty, i.e. a deviation of 10\% in each fitted value parameter (red band).  The  purple and red bands in the bottom panel correspond to the bands with the same colors on the top panel, for $E_{SNM}/A$. The uncertainties on $E_{SNM}/A$ have a quite strong effect above 2.5-3$\rho_0$. 
The precise extraction of $S(\rho)$ at high densities is inaccessible
because the high density behaviour of $E_{SNM}(\rho)/A$ is still unknown. \\

\begin{table*}[!t]
    \centering
    \begin{tabular}{cccccccccccc}
    \hline
 $E_{SNM}$ & $\rho_{max} $      & ${E'}_{0}$& ${K'}_0 $& ${J'}_{0} $& ${Z'}_{0} $ & $ {X^{(5)'}}_{0}$ & ${X^{(6)'}}_{0} $ \\
 \hline
 ${\cal O}(2)$ &$0.8$  & $-12.78\pm 0.49$ & $260.0\pm 1.1$ & $-$ & $-$ & $-$ & $-$\\
 ${\cal O}(4)$ &$0.8$  & $-16.52\pm 0.06$ & $293.4\pm 1.8$ & $55.4\pm 9.2$ & $-391\pm 16$ & $-$ & $-$\\
 ${\cal O}(6)$ &$0.8$  & $-16.05\pm 0.03$ & $242.9\pm 2.6$ & $489.5\pm 31.5$ & $-1959\pm 196$ & $1918\pm 674$ & $1488\pm 1022$\\
 ${\cal O}(2)$ &$0.32$ & $-16.21\pm 0.03$ & $280.6\pm 0.8$ & $-$ & $-$ & $-$ & $-$\\
 ${\cal O}(4)$ &$0.32$ & $-16.06\pm 0.01$ & $254.9\pm 0.8$ & $111.8\pm 10.7$ & $1011\pm 84$ & $-$ & $-$\\
 ${\cal O}(6)$ &$0.32$ & $-16.03\pm 0.001$ & $242.4\pm 0.2$ & $140.4\pm 2.1$ & $5415\pm 70$ & $-64584\pm 1890$ & $248995\pm 14940$\\
  DD2 & & $-16.02$ & $242$ & $169$ & $5234$ & $-73389$ & $483179$ \\
  \\
  \hline
  $S$ & $\rho_{max} $& ${E'}_{sym}$&${L'}_{sym}$& ${K'}_{sym} $& ${J'}_{sym} $& ${Z'}_{sym} $&  \\
  \hline
 ${\cal O}(2)$ & $0.8 $  &$ 31.54\pm 0.12$ & $47.35\pm0.42$ & $-28.7\pm 0.6$ & $-$&$-$\\
 ${\cal O}(4)$ & $0.8 $  &$ 31.53\pm 0.01$ & $56.46\pm0.08$ & $-78.6\pm 0.6$ & $112.0\pm 2.4$&$ -88.0\pm 3.6$\\
 ${\cal O}(2)$ & $0.32$  &$ 31.47\pm 0.029$ & $56.69\pm0.24$ & $-68.0\pm 1.7$ & $-$&$-$\\
 ${\cal O}(4)$ & $0.32$  &$ 31.67\pm 0.003$ & $55.28\pm0.04$ & $-97.7\pm 0.4$ & $512.1\pm 10.8$&$ -2357.2\pm 81.4$\\
   DD2 & & $31.66$ & $55.0$ & $-93.2$ & $598$ & $-5151$\\ 
  \hline
    \end{tabular}
    \caption{Fitting coefficients in Eq. (\ref{snm1}) using a fourth order and sixth order polynomial and Eq. (\ref{esym1}) using a fourth order polynomial obtained taking the exact DD2 EOS in the range $0.07<\rho<0.8$ fm$^{-3}$ and  $0.07<\rho<0.32$ fm$^{-3}$. The DD2 parameters are also shown for comparison. All quantities are given in MeV except $\rho_{max}$ which is given in fm$^{-3}$.} 
    \label{tab:par}
\end{table*}

\begin{figure}[!t]
\begin{tabular}{c}
	\includegraphics[width=.49\linewidth]{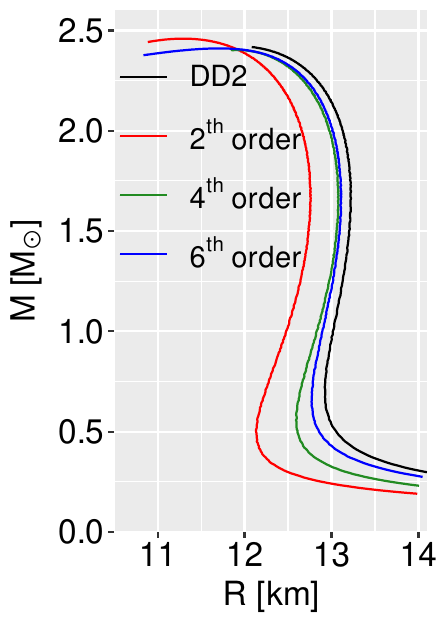} 
	\includegraphics[width=.49\linewidth]{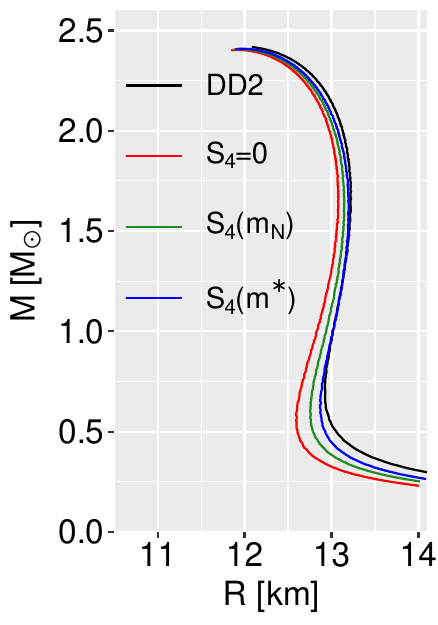} 
\end{tabular}
	\caption{Mass-radius curve for the DD2 model (full black line) and (left panel) taking the fits using Eqs. (\ref{snm1}) (fourth and sixth order) and (\ref{esym}) (fourth order), (right panel) taking the fourth order fits and including the $S_4$ term in the energy calculated with the vacuum mass or the effective mass at saturation. No muons have been included in the $\beta$-equilibrium EOS, i.e., $\varepsilon _{\beta }^{e}$.}
		\label{fig:mr}
\end{figure}

In Fig. \ref{fig:mr}, the mass-radius curves obtained integrating the TOV equations are shown for the DD2 model and two different studies. On the left panel different orders of the expansion of the EOS are considered: (i) the second order expansion of both the the symmetric nuclear matter and symmetry energy (red curve); (ii) the fourth order expansion of the symmetry energy together with   a fourth order (green curve) and a sixth order (blue curve) expansion for the symmetric nuclear matter. For stars with masses above 1$M_\odot$ the two  curves obtained using terms until 4th or 6th order in Eq. (\ref{snm}) are coincident and  very close to the $M(R)$ curve of the DD2 model: the deviation on the radius is below 2\%, see left panel. On the other hand, keeping only terms until second order has a strong effect: stars with a mass of the order of 1.5$M_\odot$ have radii $\approx 400$m smaller. This could be expected because the fit obtained with the second order expansion gives a quite small slope of the symmetry energy. It has been shown that for low mass stars there is a correlation between the slope  $L_{sym}$ and the radius, smaller radii being obtained with smaller values of $L_{sym}$, see \cite{Carriere2002,Providencia2013,Alam2016}.  The main conclusion drawn from the left panel is that the mass-radius curve is sensitive to fourth order terms but not to sixth order terms.
On the right panel, the symmetric nuclear matter and symmetry energy are described by a fourth order expansion and different approaches concerning   the fourth order contribution on the asymmetry parameter $\delta$ are considered.  
The term $S_4$ explains the deviation  of the  approximate M(R) curves  from the exact values: the exclusion of this contribution results in intermediate mass stars with $\approx 200$ m smaller radii. This could be an indication that observational data could give some information with respect to terms in the EOS having a fourth order dependence on the asymmetry parameter.
The coefficients chosen to build the approximate EOS are given in Table \ref{tab:par}. 

\section{Conclusions \label{sec4}}

In the present study we have evaluated the uncertainties associated with the determination of the symmetry energy from the $\beta$-equilibrium EOS, when a Taylor expansion is used to express the symmetric nuclear matter energy per particle and the symmetry energy. As reference we have considered the nuclear model DD2, a relativistic mean field model with density dependent couplings.

Expressing the proton fraction in terms of the $\beta$-equilibrium EOS,  the electron energy density and chemical potential and the symmetric nuclear matter energy per particle, several approximations were tested, in particular, the uncertainties due to not considering the onset of muons and the ones due to expressing the EOS in terms of a Taylor expansion until different orders. It was shown that when no muons are considered the proton fraction comes out about 7\% smaller than the exact value at densities close to the saturation density. Allowing for the onset of muons in the $\beta$-equilibrium EOS, the uncertainty on the proton fraction increases to 10\% to 15\% close to saturation and above muon onset. If, however, in the analysis to extract the proton fraction the contribution of muons is accounted for the uncertainty falls back to about 7\%.  This 7\% uncertainty was attributed to the higher order terms on the asymmetry $\delta$, i.e terms of fourth order or above that in the DD2 model arise from the kinetic energy density.

We have next extracted the symmetry energy from the proton fraction, allowing for a large uncertainty on the symmetric nuclear matter energy per particle. It was shown that while the symmetry energy is well reproduced with a fourth order density expansion function, this is not true for the symmetric nuclear matter energy. In this case the Taylor expansion fails already below 2$\rho_0$. However, it is possible to reproduce the DD2 EOS taking the Taylor expansion as a functional whose parameters can be fitted to the EOS. In this case, the parameters do not coincide with the corresponding nuclear matter properties at saturation, except for the lower order parameters if a large enough number of terms is considered in the expansion, or if the fit is taken in a  range of densities not too far from the saturation density.

It was shown that it is possible to recover the symmetry energy at saturation and below within 10\% uncertainty expressing the symmetric nuclear matter EOS as a quadratic function of the density and taking for the expansion parameters the ones obtained from chiral EFT approaches. 

Finally, it was also shown that a fourth order expansion of the symmetry energy and symmetric nuclear matter energy is enough to reproduce the mass-radius DD2 curve above 1$M_\odot$. A small difference is obtained that can be easily attributed to the lack of the higher order terms in the asymmetry parameter $\delta$.

The main conclusions from the present work can be summarized as: (i) observations could be sensitive to the asymmetry parameter dependence of the EOS. Making the parabolic approximation has two main effects: the symmetry energy is larger, since it incorporates  all the higher order terms in the asymmetry parameter that are positive, and it gives rise to smaller proton fractions inside $\beta$-equilibrium matter, since neglecting the $S_4$ term in Eq. (\ref{final2}) is equivalent to defining an effective larger asymmetry parameter; (ii) the extraction of the symmetry energy is affected by the chemical composition considered. The presence or not of muons  affects the extraction of the proton fraction above the muon onset density, a  sub-saturation density; (iii) some of these effects may mask the others. Extracting  the symmetry energy from the $\beta$-equilibrium equation of state, as done for instance in \cite{Essick2021a,Essick2021b},  may carry two types of uncertainties:  an overestimation of the symmetry energy if the terms above the parabolic approximation are neglected;  an underestimation of the proton fraction, and, therefore, of the symmetry energy, if the muon contribution is not taken into account. Finally it was also shown that   the mass-radius curve is sensitive to  the fourth order terms of the EOS but not to sixth order terms and, besides, it is also sensitive to the terms on the asymmetry parameter above the parabolic approximation.

The determination of the composition of a NS from the $\beta$-equilibrium  EOS is a challenge that requires the knowledge of the symmetry energy. In the present study we have only considered protons, neutrons, electrons and muons but at high densities other species as hyperons, deltas, boson condensates or quark phases may also occur, which rise further difficulties.

\section{Acknowledgments}

This work was partially supported by national funds from FCT (Fundação para a Ciência e a Tecnologia, I.P, Portugal) under the Projects No. UIDB/FIS/04564/2020, No. UIDP/04564/2020, and No. POCI-01-0145-FEDER-029912 with financial support from Science, Technology and Innovation, 
in its FEDER component, and by the FCT/MCTES budget through national funds (OE).


%

\section{Appendix}

The sixth order $S_6(\rho)$ and eight order $S_8(\rho)$ terms of the expansion of the energy per particle in the asymmetry parameter $\delta$
\begin{widetext}
\begin{equation}S_{6}\left ( \rho  \right )=\frac{1}{6!3^{5}}\frac{880\,k_{F}^{10}+2140\,k_{F}^{8}m^{*2}+2385\,k_{F}^{6}m^{*4}+1300\,k_{F}^{4}m^{*6}+280\,k_{F}^{2}m^{*8}}{\epsilon_F^{9}},
\label{s6}
\end{equation}
and the eighth order
\begin{equation}S_{8}\left ( \rho  \right )=\frac{35}{8!3^{7}}\frac{\left (5984\,k_{F}^{14}+23088\,k_{F}^{12}m^{*2}+43578\,k_{F}^{10}m^{*4}+47737\,k_{F}^{8}m^{*6}+30888\,k_{F}^{6}m^{*8}+10992\,k_{F}^{4}m^{*10}+1664\,k_{F}^{2}m^{*12}  \right )}{\epsilon_F^{13}}.
\label{s8}
\end{equation}

\end{widetext}
\end{document}